# Microscopic/stochastic timesteppers and "coarse" control: a kinetic Monte Carlo example.

C. I. Siettos[1], A. Armaou[1], A. G. Makeev[1,#] and I.G. Kevrekidis[1*]

[1]Department of Chemical Engineering, Princeton University, Princeton, NJ 08544, USA

[#]Permanent Address: Faculty of Computational Mathematics and Cybernetics,

Moscow State University, Moscow, 119899, Russia

*Keywords*: timesteppers, system identification, kinetic Monte Carlo, multiscale computation

**Abstract:** Coarse timesteppers provide a bridge between microscopic / stochastic system descriptions and macroscopic tasks such as coarse stability/bifurcation computations. Exploiting this computational enabling technology, we present a framework for designing observers and controllers based on microscopic simulations, that can be used for their *coarse control*. The proposed methodology provides a bridge between traditional numerical analysis and control theory on the one hand and microscopic simulation on the other.

**Introduction.**   Mathematical models, whether identified on-line during an experiment or derived from first principles, constitute the backbone of modern control practice. Models of chemical and transport processes (material, energy and momentum balances) traditionally take the form of deterministic ordinary or partial differential/algebraic evolution equations for macroscopic variables (e.g. concentrations). An arsenal of mathematical and computational tools targeted at such macroscopic models has been developed over the years for the performance of macroscopic, system-level tasks such as temporal simulation, stability and bifurcation analysis, optimization, design and control. For many processes of current interest, however, the best available description of the physics (through molecular dynamics, MD, kinetic Monte Carlo, KMC, kinetic theory based Lattice-Boltzmann, LB, or Markov chain simulators) operates at a vastly different scale from that at which the questions of interest are asked and the answers are required (e.g. operating diagrams and controller design for expected reaction rates). The implication is that macroscopic rules (description at a higher level) can somehow be deduced from microscopic ones (description at a much finer level). In most current problems, however, ranging from ecology to materials science and from chemistry to engineering, the closures required to translate microscopic/stochastic models to a high-level, macroscopic description are simply not available. A so-called "coarse timestepper" approach to stability and bifurcation calculations has recently had some success in circumventing the derivation (through explicit closures) of macroscopic descriptions (Theodoropoulos et al., 2000, Gear et al., 2002, Makeev et al., 2002, Runborg et al., 2002). Using short "bursts" of appropriately initialized and processed microscopic simulations coupled with system identification, one tries to analyze the behavior of macroscopic equations *without ever obtaining these equations in closed form.*

---

[*]To whom correspondence should be addressed

The purpose of this letter is to suggest that such "coarse timesteppers" have the potential to become a bridge between microscopic/stochastic modeling and well-established control design techniques, such as linear feedback or model predictive control. The key assumption is that deterministic, macroscopic, *coarse* models exist *and close* for the expected behavior of a few low *moments* of microscopically evolving distributions (e.g. for surface coverages, the zeroth moments of adspecies distributions on a lattice); but they are unavailable in closed form. Computer-assisted location of steady states or design of simple stabilizing controllers requires the *evaluation* of certain quantities from these unavailable equations (residuals, Jacobians, Hessians etc.). One formulates a framework in which it is possible to computationally *identify* these quantities "*on demand"* (or *"just in time",* (Cybenko, 1996)) from judicious microscopic simulations, since, lacking the explicit equation, we cannot simply evaluate them. We then pass them to the algorithm that performs the macroscopic task (steady state location, control design algorithm). In effect, through a computational superstructure built around the microscopic simulation code, we "fool" the macroscopic numerical algorithms into performing tasks for an equation that is not explicitly available.

We start with a brief overview of the "coarse timestepper" (detailed discussions can be found in Gear et al., 2002, Makeev et al., 2002), and a simple version of the framework for macroscopic controller design based on microscopic simulators. Our illustrative example is a kinetic Monte Carlo (KMC) simulation of a simple surface reaction scheme. We first obtain its timestepper-based coarse bifurcation diagram. We then demonstrate the results of implementing a simple *coarse controller* for stabilizing an open-loop unstable coarse (macroscopic) equilibrium of the KMC model, and conclude with a short discussion.

**Coarse controllers for microscopic simulators.** The following steps outline a simple framework for designing coarse, macroscopic controllers, based on information extracted at the system level from microscopic models, sidestepping the derivation of a closed form macroscopic description:

    **1. The Coarse Time-Stepper**

**(a)** Choose appropriate statistics **x** (typically zeroth- or first-order moments of evolving probability densities **X**) that we believe can deterministically describe the long-term *macroscopic* behavior of the system under study. This implies that higher-order statistics become quickly (say, over a few collision times in a microscopic simulation) slaved to these lower ones (they evolve to functionals of the lower ones, alternatively evolve towards a "slow manifold" parametrized by the lower ones). These choices also determine a *restriction* operator, $M$, mapping (projecting) the microscopic-level description **X**, into the macroscopic description: $\mathbf{x} = M\mathbf{X}$;

**(b)** Choose an appropriate (non-unique !) *lifting* operator, $\mu$, mapping the macroscopic description, **x**, to a consistent microscopic description, **X**; that is, construct distribution(s) consistent with *(conditioned on)* a few low-order moments (see Gear et al., 2002, Makeev et al., 2002 for examples and discussion). Lifting from the microscopic to the macroscopic and then restricting again should have no effect, that is, $M\mu = I$ (except roundoff);

**(c)** Prescribe a macroscopic initial condition (e.g. concentration profile, coverage) $\mathbf{x}(t_0)$;

**(d)** Transform it through lifting to one (or more) consistent microscopic realizations $\mathbf{X}(t_0) = \mu \mathbf{x}(t_0)$;

**(e)** Evolve this(ese) realization(s) using the microscopic simulator for the desired short macroscopic time T, generating the value(s) $\mathbf{X}(T)$. The choice of T is associated with the (estimated) spectral gap of the linearization of the unavailable macroscopic equation, (see Gear et al., 2002, Makeev et al., 2002 for discussion);

**(f)** Obtain the restrictions $\mathbf{x}(T) = M\mathbf{X}(T)$.

The procedure is a "black box" "coarse timestepper" $\Phi_T(\mathbf{x}_0) = \mathbf{x}(T)$ with $\mathbf{x}_0$ as initial condition.

**2. A Coarse Stability/Bifurcation Framework** can now be implemented as a computational superstructure, a "shell", around repeated calls to this coarse timestepper. *Coarse (macroscopic) steady states* can be obtained as fixed points of the iteration $\mathbf{x}(k+1) = \Phi_T(\mathbf{x}(k))$. These are **not** steady states of the (stochastic, constantly varying) KMC simulation; they are "almost always" steady states of the (expected values of the) low moments **x** of the KMC-evolved distributions **X**. The action of the (slow) local Jacobian of the mapping, $D\Phi$, is estimated "on demand" using numerical derivatives and "lift-run-restrict" evaluation of $\Phi$. For large, discretized distributed coarse systems, calls to the timestepper with $\boldsymbol{\varepsilon}$-nearby coarse initial conditions are used to estimate the requisite matrix-vector products $\mathbf{D\Phi}\cdot\boldsymbol{\varepsilon}$. Contraction mappings like Newton-Raphson, or so-called Newton-Picard timestepper based algorithms using Krylov subspace iterations (e.g. the Recursive Projection Method of Shroff and Keller, 1993) are used to find the coarse steady states (see also Lust, 1997, Tuckerman and Barkley, 2000).

Coarse input-output data, especially in the neighborhood of the located macroscopic steady state, can be used to perform additional coarse (e.g. nonlinear ARMAX) model identification; extended Kalman filters, extended least squares (ELS) or recursive maximum likelihood (ML) algorithms (Åstrom and Wittenmark, 1995, Ljung, 1999) can be used for the estimation of the polynomial coefficients. The important element is that the initial conditions of the coarse time-stepper (as opposed to those of a physical experiment) can be repeatedly set *at will*. We can now proceed to:

**3. A Coarse Control Framework**. The above computational framework serves as an "on demand" identification methodology for right-hand-sides, "coarse slow" Jacobians, coarse derivatives with respect to parameters etc.; in short, of precisely the quantities that a control design algorithm would need evaluated from a macroscopic model, had such a model been available, to perform its task. We simply substitute the macroscopic function evaluations with this computational shell around the microscopic timestepper. Using the concepts of separation and certainty equivalence principles we then perform control tasks, such as the design of local linear stabilizing controllers. Optimal controller and model predictive controller design, even though considerably more intricate, conceptually follow directly in a discrete-time, coarse timestepper based framework.

**A simple kinetic Monte Carlo model for catalytic CO oxidation.** Our illustrative model is a "stochastic simulation algorithm" (Gillespie, 1976 and 1977) Monte Carlo realization of a simplification of the kinetics of catalytic CO oxidation whose mean field description is given by the following equations:

$$\frac{d\theta_A}{dt} = a\theta^* - \gamma\theta_A - 4k_r\theta_A\theta_B \tag{1a}$$

$$\frac{d\theta_B}{dt} = 2\beta\theta^{*2} - 4k_r\theta_A\theta_B \tag{1b}$$

$$\frac{d\theta_C}{dt} = \mu\theta^* - \eta\theta_C \tag{1c}$$

where $\theta_i$ represent the coverages of species (i=A,B,C, resp. CO, O and an inert species C) on the catalytic surface; $\theta^* = 1 - \theta_A - \theta_B - \theta_C$; $\alpha,\beta,\gamma,\mu,\eta$ are associated with CO adsorption, O dissociative adsorption, CO desorption, and C ads/desorption rates; and $k_r$ is the reaction rate constant. Simulation results were obtained for $\alpha = 0.01$, $\gamma = 0.04$ and $k_r =1$, $\eta = 0.016$. The bifurcation parameter was $\beta$ and the control variable was $\mu$ (physically varied through gas phase pressures of $O_2$ and the inert C respectively). The KMC simulations approximate the solution of the corresponding master equation, which describes the evolution of the probability (PDF) of finding the system in a certain configuration (Makeev et al., 2002). Figure 1a shows the open loop behavior (at $\beta \approx 20$ and $\mu =0.36$) of $\theta_A$ using the KMC simulator for two system sizes, $N=200^2$ and $N=1000^2$, and a single realization $N_{run}=1$; the deterministic response is also given for comparison purposes. Fig. 1b shows a one-parameter bifurcation diagram (both mean-field and the coarse KMC timestepper one) with respect to $\beta$. The model exhibits two supercritical Hopf bifurcations (at $\beta \approx 20.3$ and $\beta \approx 21.2$) and stable oscillatory behavior in-between.

We want a stabilizing controller for the macroscopic (expected) *unstable* steady state at $\beta \approx 20.6879$. This coarse steady state (evaluated through the T=0.025 coarse timestepper, upon convergence of the Newton-Raphson to a residual of $O(10^{-5})$ for $\varepsilon \sim 10^{-2}$) is $\theta_A \approx 0.294092$, $\theta_B \approx 0.029173$, $\theta_C \approx 0.648099$. At stationarity, the estimates of the Jacobian **A** and control matrix **B** of the unknown macroscopic equation are

$$\mathbf{A} \approx \begin{bmatrix} 0.965136 & -0.065787 & -0.034237 \\ -0.051537 & 0.920113 & -0.05825 \\ -0.012137 & -0.009963 & 0.988712 \end{bmatrix}, \mathbf{B} \approx \begin{bmatrix} -1.637E-04 \\ -3.475E-04 \\ 7.475E-04 \end{bmatrix};$$

(This coarse steady state is estimated within 1% of the mean field one; the Jacobian and control matrix elements are also well estimated). These matrices, which would appear in a standard discrete time local linear stochastic state space model of the type $\mathbf{x}(k+1) = \mathbf{A}\mathbf{x}(k) + \mathbf{B}\mathbf{u}(k) + \mathbf{w}(k)$, (where the vector $\mathbf{x} \in \tilde{N}^n$ represents the state variables; $\mathbf{u} \in \tilde{N}^p$ is the control vector of system while the vector $\mathbf{w} \in \tilde{N}^n$ denotes the process noise), are a byproduct of the fixed point/ continuation algorithm for coarse steady state location. For large-size problems (such as those arising in discretized coarse PDEs), RPM-type algorithms will identify the "coarse slow" Jacobian (for dissipative PDEs in which a separation of "coarse slow" and "coarse fast" time scales arises naturally). Such estimates can be made more accurate using the coarse timestepper as an experiment, repeatedly initializing it at will in the neighborhood of the coarse steady state. Discrete time, stochastic state-space models of the standard form above, along with Gaussian uncorrelated noise sequence assumptions for $\mathbf{w}(k)$, bring textbook linear identification theory to bear on the problem; colored noise or even nonlinear identification tools may be progressively incorporated.

For our linear feedback controller design we consider the case of high process noise, dictating an observer for the estimation $\hat{\mathbf{x}}$ of the deviation state variables from the coarse steady state. The controller thus takes the form of a linear feedback $\mu - 0.36 \equiv u = -\mathbf{K}\hat{\mathbf{x}}$ where the gain matrix $\mathbf{K}$ is calculated on the basis of the separation principle The simple illustrative choice here is to use a discrete Kalman filter (Kalman and Bucy, 1961, Sage and White, 1980) as an observer, and pole placement (Kailath, 1980) for the controller design. The eigenvalues of the *discrete time* open loop system as calculated from the estimated coarse Jacobian are $\lambda_1 = 0.87135$, $\lambda_{2,3} = 1.00130 \pm 0.00529i$, consistent with instability and oscillatory behavior. The noise covariance matrix, with covariance $\text{Cov}[\mathbf{w}(k), \mathbf{w}(j)] = \mathbf{W}\,\delta(k-j)$, required for Kalman filter design, was obtained numerically from the microscopic simulator. To stabilize the unstable coarse steady state, we placed the coarse eigenvalues to: $\lambda_1 = 0.87135$, $\lambda_2 = 0.98$ and $\lambda_3 = 0.99$. The required gains were $k_1 \approx 32.52$, $k_2 \approx -28.55$, $k_3 \approx 23.23$. The result of these choices is shown in Fig.2a, comparing open-loop and closed-loop responses of $\theta_A$ for the KMC model with $N=200^2$, and $N_{run} =1$. One can claim that this simple control scheme, using a discrete Kalman filter as an observer, locally stabilizes the unstable coarse steady state of the (expected value of the) system. Figure 2b shows he corresponding dynamics of the control variable. Under comparable considerations we have promising results (Rico-Martinez, 2001, Siettos et al., 2002) for using coarse control as a part of an adaptive scheme that enables microscopic simulators (or experiments) to automatically trace coarse bifurcation diagrams and to converge to low codimension coarse bifurcation points. Variance reduction plays a vital role in all these tasks.

**Discussion**. The proposed framework aims at establishing a synergism between "conventional" control design techniques on the one hand, and microscopic complex systems modeling on the other. Let us make clear that this procedure can only be useful if a macroscopic description is *conceptually possible* yet unavailable in closed form; if accurate macroscopic models *are* available in closed form, one of course should use them for designing model-based controllers directly. What is proposed here, is a systematic way of designing controllers for microscopic models based on the information one would obtain from macroscopic models, had these models been available in closed form. C*onceptual* extensions to coarse time-stepper based model-predictive control, (local) feedback-linearization, optimal control, and "coarse optimization" appear straightforward, although there will be many nontrivial theoretical and implementation problems. System based techniques (identification) and ideas (separation of time scales between "governing" and "slaved" moments for the microscopic dynamics, as well as between "slow coarse" and "fast coarse" modes in dissipative PDEs) lie at the core of this "computational enabling technology" that bridges microscopic timesteppers with system-level tasks. This work was partially supported by AFOSR (Dynamics and Control, Drs. Jacobs and King) and the National Science Foundation.

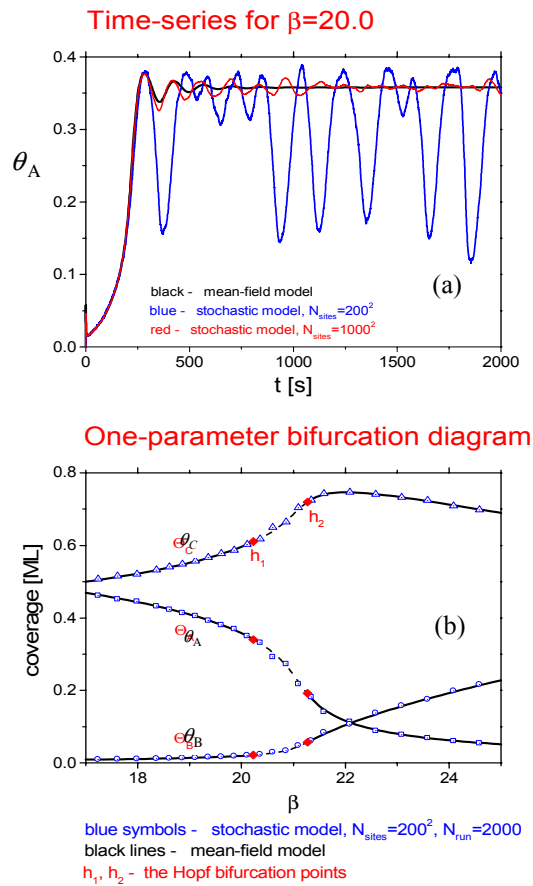

Figure 1. (a) Open loop $\theta_A$ response (KMC simulator, $\beta \approx 20$ and $\mu = 0.36$, system sizes $N=200^2$ (blue line) and $N=1000^2$ (red line), and a single realization $N_{run}=1$); the deterministic response (black line) is also given for comparison purposes. (b) One-parameter coarse KMC bifurcation diagram. Two supercritical coarse Hopf bifurcation points are marked with diamonds; circles, squares and triangles correspond to the coarse KMC steady states. Mean field bifurcation diagram (for comparison); stable(unstable) steady states are represented by solid (dotted) lines.

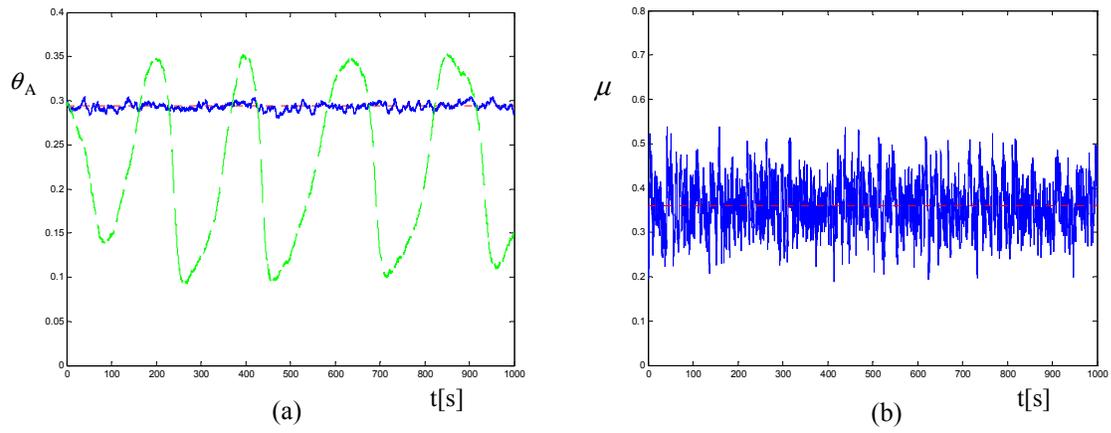

Figure 2. (a) Open loop (dashed line) and closed loop (blue solid line) $\theta_A$ response (KMC simulator, $N=200^2$, $N_{run} = 1$); (b) corresponding closed loop response of the manipulated variable $\mu$ (blue solid line); red dotted lines correspond to nominal steady state conditions.

9